\documentclass{article}
\usepackage[psamsfonts]{amssymb}
\usepackage{amsmath}
\usepackage{cite}

\author {Mohammad~Khorrami$^1$\footnote{mamwad@mailaps.org} \&
Masoud~Alimohammadi$^2$\footnote{alimohmd@ut.ac.ir}
\\ $^1$ {\small Department of Physics, Alzahra University, Tehran
1993891167, Iran.}
\\ $^2$ {\small Department of Physics, University of Tehran,}
\\ {\small North Karegar Ave., Tehran, Iran.}}
\title{Non-Douglas-Kazakov phase transition of two-dimensional generalized Yang-Mills theories}
\date{}
\begin{document}
\maketitle
\begin{abstract}
\noindent In two-dimensional Yang-Mills and generalized Yang-Mills
theories for large gauge groups, there is a dominant
representation determining the thermodynamic limit of the system.
This representation is characterized by a density the value of
which should everywhere be between zero and one. This density
itself is determined through a saddle-point analysis. For some
values of the parameter space, this density exceeds one in some
places. So one should modify it to obtain an acceptable density.
This leads to the well-known Douglas-Kazakov phase transition. In
generalized Yang-Mills theories, there are also regions in the
parameter space where somewhere this density becomes negative.
Here too, one should modify the density so that it remains
nonnegative. This leads to another phase transition, different
from the Douglas-Kazakov one. Here the general structure of this
phase transition is studied, and it is shown that the order of
this transition is typically three. Using carefully-chosen
parameters, however, it is possible to construct models with
phase-transition orders not equal to three. A class of these
non-typical models are also studied.
\end{abstract}
\newpage
\section{Introduction}
The two-dimensional Yang-Mills theory (YM$_2$) is a laboratory for
testing ideas and concepts about the Yang-Mills model in the real
four dimensional world. The string picture of YM$_2$ is also
interesting on its own, as an example for the nonperturbative
analysis of a quantum field theory \cite{1,2,3}.

The starting point to make a correspondence between YM$_2$ and
string theory, is to study the large--$N$ limit of YM$_2$. For
example, as it is shown in \cite{2}, a gauge theory based on
SU($N$) is split at large $N$ into two copies of a chiral theory,
which encapsulate the geometry of the string maps. The chiral
theory associated with the Yang--Mills theory on a two--manifold
$\cal M$ is a summation over maps from a two--dimensional world
sheet (of arbitrary genus) to the manifold $\cal M$. This leads to
a $1/N$ expansion for the partition function and observables that
is convergent for all of the values of area$\times$coupling
constant on the target space $\cal M$, if the genus is one or
greater.

The large-$N$ limit of the U$(N)$ YM$_2$ on a sphere has been
studied in \cite{4}. There the dominant (or classical)
representation has been found and it has been shown that the free
energy of the U$(N)$ YM$_2$ on a sphere with the surface area
$A<A_{\mathrm{c}}=\pi^2$ has a logarithmic behavior . In \cite{5},
the free energy has been calculated for areas $A>\pi^2$, from
which it has been shown that the YM$_2$ on a sphere has a
third-order phase transition at the critical area
$A_{\mathrm{c}}=\pi^2$. This is the well-known Douglas-Kazakov
phase transition. It resembles the Gross-Witten-Wadia phase
transition for the lattice two dimensional multicolour gauge
theory \cite{6}.

The main characteristics of YM$_2$ are its invariance under
area-preserving diffeomorphisms and the fact that there are no
propagating degrees of freedom. These properties are not unique to
YM$_2$, but rather are shared by a wide class of theories, called
the generalized Yang-Mills theories (gYM$_2$'s) \cite{7}. Various
properties of gYM$_2$'s have been studied, including their phase
structure \cite{8}, the large-$N$ behavior of the Wilson loops
\cite{9}, and their double-scaling limit properties \cite{10}.

The phase structure of gYM$_2$'s is an interesting issue from
various points of view. The reason the Douglas-Kazakov (DK) phase
transition occurs in YM$_2$ and gYM$_2$'s, is that in $A<A_c$
areas (the so called the weak regime) the dominant density
function determined through a saddle-point analysis
($\rho_{\mathrm{w}}$) is everywhere less than one. But for $A>A_c$
areas (the strong regime) $\rho_{\mathrm{w}}$ becomes greater than
one in some points, which is not acceptable and so
$\rho_{\mathrm{w}}$ must be replaced by a new density
($\rho_{\mathrm{s}}$). The transition from $\rho_{\mathrm{w}}$ to
$\rho_{\mathrm{s}}$ in YM$_2$ induces a phase transition of order
three. The DK phase transition has a richer structure in gYM$_2$'s
and the order of which may be other than three, in the so called
nontypical theories \cite{8}.

Another feature of the phase structure of gYM$_2$'s, which does
not exist in YM$_2$, is the possibility that $\rho_{\mathrm{w}}$
becomes negative somewhere. A negative density is not acceptable
either, so it must be replaced by another density, which we denote
it by $\rho_{\mathrm{g}}$ (``g'' for the ``gaped phase''), in
which the density function is zero in some region. In \cite{11}, a
special gYM$_2$ has been studied, in which the action is a
combination of quartic and quadratic Casimirs,  and a third order
phase transition from the weak regime to the gaped regime has been
observed. This observation is based on a classification done by
Jurekiewicz and Zalewski \cite{12}.

In the present work, the structure of the transition between the
weak regime and the gaped regime for arbitrary gYM$_2$'s is
studied. The technique used is similar to that used in \cite{8}
and \cite{9} to study the DK phase structure of gYM$_2$'s. In this
method, only the behavior of $\rho_{\mathrm{w}}$ near the critical
point is needed. This is important, since the functional form of
$\rho_{\mathrm{w}}$ is explicitly known for all gYM$_2$'s, while
that's not the case for $\rho_{\mathrm{s}}$ and
$\rho_{\mathrm{g}}$. It is shown that the non-DK phase transitions
between the weak and gaped regimes are in \textit{almost} all
gYM$_2$'s of order three. There are, however, some special cases
with fine-tuned couplings where one can have the phase transition
of nontypical orders.

The scheme of the paper is the following. In section 2, gYM$_2$'s
are briefly reviewed and the method of calculating the order of
non-DK phase transitions is discussed. In section 3, this method
is applied to the typical model discussed in \cite{11} and it is
proved that the order of this non-Dk phase transition is three.
Finally, in section 4 a nontypical gYM$_2$ is introduced the order
of its weak-gaped transition is 5/2 and an outline is given on how
to construct a general nontypical model with the non-DK phase
transition of order $2+(1/k)$, where $k$ is an integer greater
than one.

\section{The general method}
The partition function of the generalized U$(N)$ Yang-Mills theory
on a sphere of area $A$ is \cite{7,13}
\begin{equation}\label{1}
Z=\sum_rd_r^2\,e^{-A\,\Lambda (r)},
\end{equation}
where $r$'s label the irreducible representations of the group
U$(N)$, $d_r$ is the dimension of the representation $r$, and
\begin{equation}\label{2}
\Lambda(r)=\sum_{k=1}^p\frac{a_k}{N^{k-1}}\,C_k(r).
\end{equation}
$C_k$ is the $k$'th Casimir of the gauge group, and $a_k$'s are
arbitrary constants. For the partition function (1) to be
convergent, it is necessary that $p$ in (2) be even and $a_p$ be
positive. The representations of $U(N)$ are parameterized by
integers $n_i$ such that $n_1\geq n_2\geq\cdots\geq n_N$.

In the  large-$N$ limit, it is convenient to introduce the
continuous variable
\begin{equation}\label{3}
\phi(x):=-n(x)-1+x,
\end{equation}
where
\begin{align}\label{4}
0 \leq x:=&\frac{i}{N}\leq 1,\nonumber\\
n(x):=&\frac{n_i}{N}.
\end{align}
The partition function (\ref{1}) then becomes
\begin{equation}\label{5}
Z=\int\prod_{0\leq x\leq 1}\mathrm{d}\phi(x)\;e^{S(\phi)},
\end{equation}
where
\begin{equation}\label{6}
S(\phi):=N^2\left\{-A\,\int^1_0\mathrm{d}x\;
G[\phi(x)]+\int_0^1\mathrm{d}x\int_0^1\mathrm{d}y\;\log|\phi(x)-
\phi(y)|\right\},
\end{equation}
(apart from an unimportant constant), and
\begin{equation}\label{7}
G(\phi ):=\sum_{k=1}^p(-1)^k\,a_k\,\phi^k.
\end{equation}
Introducing the density function
\begin{equation}\label{8}
\rho[\phi(x)]:=\frac{\mathrm{d}x}{\mathrm{d}\phi(x)},
\end{equation}
it is seen that it satisfies
\begin{equation}\label{9}
\int^a_{-a}\mathrm{d}z\;\rho(z)=1,
\end{equation}
where $[-a,a]$ is the interval corresponding to the values of
$\phi$. Here it is assumed that $G(\phi)$ is even, and therefore
$\rho(z)$ is even as well. The condition $n_1\geq
n_2\geq\cdots\geq n_N$ demands
 \begin{equation}\label{10}
 0\leq \rho (z)\leq 1.
 \end{equation}
As $N$ tends to infinity, the only representation that contributes
to the partition function (\ref{5}) is the so called dominant (or
classic) representation \cite{14}, satisfying
\begin{equation}\label{11}
g(z)=\mathrm{P}\int^a_{-a}\mathrm{d}z'\;\frac{\rho(z')}{z-z'},
\end{equation}
where $\mathrm{P}$ indicates the principal value of the integral,
and
\begin{equation}\label{12}
 g(z):=\frac{A}{2}\,G'(z).
\end{equation}
The free energy of the theory is defined through
\begin{equation}\label{13}
F:=-\frac{1}{N^2}\ln Z.
\end{equation}
Using the standard method of solving the integral equation
(\ref{11}), the density function $\rho$ is obtained, following
\cite{14}, as
\begin{equation}\label{14}
\rho(z)=\frac{\sqrt{a^2-z^2}}{\pi}\,\sum_{n,q=0}^\infty
\frac{(2n-1)!!}{2^n\,n!\,(2\,n+2\,q+1)!}\,a^{2\,n}
\,z^{2\,q}\,g^{(2\,n+2\,q+1)}(0),
\end{equation}
where the parameter $a$ satisfies
\begin{equation}\label{15}
\sum_{n=1}^\infty\frac{(2\,n-1)!!}{
2^n\,n!\,(2n-1)!}\,a^{2\,n}\,g^{(2\,n-1)}(0)=1.
\end{equation}
Here $g^{(n)}$ is the $n$-th derivative of $g$.

The density function in (\ref{14}), which we call
$\rho_{\mathrm{w}}$, should satisfy the conditions (\ref{10}). It
depends obviously on the area $A$ and the parameters $a_k$, and
there could be regions in the parameter space (of $A$ and $a_k$'s)
where these conditions are violated. In such regions, the density
corresponding to the dominant representation is not
$\rho_{\mathrm{w}}$. The case where $\rho_{\mathrm{w}}(z)$ exceeds
one for some $z$ results in the well-known DK phase transition.
Here we are interested in the case where $\rho_{\mathrm{w}}(z)$
becomes negative for some $z$. In this case $\rho_{\mathrm{g}}$,
the density corresponding to the dominant representation, is zero
in some interval $L_{\mathrm{g}}$, the length of which is defined
$2\,b$. For the same parameters, $\rho_{\mathrm{w}}$ is negative
in a region $L'_{\mathrm{g}}$, the length of which is of the order
$2\,b$. Defining
\begin{equation}\label{16}
\alpha:=|\min(\rho_\mathrm{w})|,
\end{equation}
one can use exactly the same arguments used in \cite{9} for the DK
phase transition to show that
\begin{equation}\label{17}
H_\mathrm{g}(z)-H_\mathrm{w}(z)\sim
b^{2\,m+2}\sim\alpha^{1+(1/m)}, \qquad\hbox{for large } z.
\end{equation}
Here
\begin{equation}\label{18}
H_{\mathrm{g},\mathrm{w}}(z):=\int\mathrm{d}y\;
\frac{\rho_{\mathrm{g},\mathrm{w}}(y)}{{z-y}},
\end{equation}
$2\,m$ is the order of the first nonvanishing derivative of
$\rho_\mathrm{w}$ at the point $\rho_\mathrm{w}$ attains its
minimum, and by large $z$ it is meant $|z|>>a$.

Assume that the phase transition from $\rho_\mathrm{w}$ to
$\rho_\mathrm{g}$ occurs at some critical value $A=A_\mathrm{c}$.
If the order of the first nonvanishing derivative of $\alpha$ with
respect to $A$ at the point $A_\mathrm{c}$ is $l$, then (\ref{17})
results in
\begin{equation}\label{19}
H_\mathrm{g}-H_\mathrm{w}\sim(A -A_\mathrm{c})^{l\,[1+(1/m)]}.
\end{equation}
Using (\ref{13}), and the fact that the dominant representation
maximizes $S$, it is seen that
\begin{align}\label{20}
F'(A)&=\int_0^1\mathrm{d}x\;G[\phi(x)],\nonumber\\
&=\int\mathrm{d}y\;\rho(y)\,G(y),\nonumber\\
&=\frac{1}{2\,\pi\,i}\oint\mathrm{d}z\;H(z)\,G(z),
\end{align}
where the last integration is over a large contour and in the last
step (\ref{18}) has been used. So one arrives at
\begin{equation}\label{21}
F'_\mathrm{g}(A)-F'_\mathrm{w}(A)\sim(A-A_\mathrm{c})^{l\,[1+(1/m)]},
\end{equation}
and from that
\begin{equation}\label{22}
F_\mathrm{g}(A)-F_\mathrm{w}(A)\sim(A-A_\mathrm{c})^{1+l\,[1+(1/m)]}.
\end{equation}
This is our desired relation. Eq.(\ref{22}) shows that for typical
theories where $l=m=1$, the system exhibits a third order phase
transition at $A=A_\mathrm{c}$, but for special theories (with
fine-tuned coupling constants) this order can in principle be
different from three. The situation is completely analogous to
that of the DK phase transition.

One can extend this argument to the case where several phase
transitions occur, either of them could correspond to a density
exceeding one or becoming negative. Suppose that the density is
already so that there are regions where it is identically one or
zero, and varying the parameters creates a new region where the
density either exceeds one or becomes negative. Let us call this
density $\rho_1$, and the boundary which is violated (zero or one)
$B$. This density should be corrected so that the corrected
density $\rho_2$ does not cross the boundary $B$. For the density
$\rho_2$, there is a new interval where $\rho_2$ is equal to $B$.
Defining $\alpha$ as the maximum of $|\rho_1(z)-B|$, and $2\,b$ as
the width of the region where the value of $\rho_1$ is not in
$[0,1]$, it is seen that exactly similar arguments lead to
something like (\ref{22}), where the left-hand side is
$F_2(A)-F_1(A)$. So, even if there are several transitions of
either kind (the density crossing zero or one), the order of each
transition is similar to the case of DK phase transition.
Specially, any transition which is typical ($l=m=1$) is a third
order transition.

\section{A typical model}
Consider the following typical model
\begin{equation}\label{23}
G(z)=c_2\,z^2+c_4\,z^4,
\end{equation}
which has been investigated in \cite{8,11}, where
\begin{equation}\label{24}
c_4>0.
\end{equation}
Using (\ref{14}) and (\ref{15}), one finds
\begin{align}\label{25}
\rho_\mathrm{w}(z)=&\frac{\mu}{\pi}\,\sqrt{a^2-z^2}\,
\left(z^2+\frac{a^2}{2}+\beta\right),\\ \label{26}
a^2=&-\frac{2\,\beta}{3}+\frac{2}{3}\,\sqrt{\frac{6}{\mu}+\beta^2},
\end{align}
where
\begin{align}\label{27}
\beta:=&\frac{c_2}{2\,c_4},\nonumber\\
\mu:=&2\,A\,c_4.
\end{align}
There arise three distinct regions in the $(\beta-\mu)$ plane:
\begin{itemize}
\item[\textbf{i}] $\beta>\sqrt{2/\mu}$.\\
Here $\rho_\mathrm{w}(z)$ has a maximum at $z=0$.
\item[\textbf{ii}] $-\sqrt{2/\mu}<\beta<\sqrt{2/\mu}$.\\
Here $\rho_\mathrm{w}(z)$ has a positive minimum at $z=0$ and two
maxima at $\pm z_0:=\pm\sqrt{(3\,a^2-2\,\beta)/6}$.
\item[\textbf{iii}] $\beta<-\sqrt{2/\mu}$.\\
Here $\rho_\mathrm{w}(z)$ has a negative minimum at $z=0$ and two
maxima at $\pm z_0:=\pm\sqrt{(3\,a^2-2\,\beta)/6}$.
\end{itemize}
Clearly in the region \textbf{iii}, one must replace
$\rho_\mathrm{w}$ with $\rho_\mathrm{g}$, which is equal to zero
in an even interval around $z=0$, and the system undergoes a
non-DK phase transition from the weak regime to the gaped regime
on the critical curve
\begin{equation}\label{28}
\beta=-\sqrt{\frac{2}{\mu}}.
\end{equation}
At this transition curve, the maximum value of
$\rho_\mathrm{w}(z)$ is
\begin{equation}\label{29}
\rho_\mathrm{w}(\pm
z_0)=\left(\frac{2^{13}\,\mu}{3^6\,\pi^4}\right)^{1/4}.
\end{equation}
One should also make sure that at this point $\rho_\mathrm{w}$ is
still never greater than one, which sets a condition for $\beta$:
\begin{equation}\label{30}
\beta\leq\beta_0:=-\frac{128}{27\,\pi^2}.
\end{equation}
So as $\mu$ increases, for $\beta<\beta_0$ the system goes from
the weak regime to the gaped regime, while for $\beta>\beta_0$ the
system goes from the weak regime to the strong regime, i.e.
undergoes the ordinary DK phase transition.

To obtain the order of the above non-DK phase transition, one must
determine the parameters $l$ and $m$ in (\ref{22}). Using
(\ref{25}), it is seen that
 \begin{equation}\label{31}
\rho''(0)|_\mathrm{c}=\left(\frac{128\,\mu^3}{\pi^4}\right)^{1/4},
\end{equation}
which is positive and shows that $m=1$. To obtain $l$, one should
investigate the behavior of $\alpha$ for fixed $\beta$ with
respect to $\mu$. (Note that $\mu$ differs from $A$ by just a
multiplicative positive constant, $2\,c_4$.) One has
\begin{equation}\label{32}
\alpha=-\frac{\mu\,a}{\pi}\,\left(\frac{a^2}{2}+\beta\right).
\end{equation}
From (\ref{26}), it is seen that the derivative of $a$ with
respect to $\mu$ at fixed $\beta$ is a finite negative number.
Noting that $(a^2+2\,\beta)$ vanishes at the transition, one
obtains
\begin{equation}\label{33}
\frac{\partial\alpha}{\partial\mu}=-\frac{\mu\,a^2}{\pi}\,
\frac{\partial a}{\partial\mu},
\end{equation}
which is positive. So $l=1$ as well, and the order of the
transition is 3. This has been pointed out in \cite{11}, based on
different arguments.
\section{Nontypical models}
Consider the following potential.
\begin{equation}\label{34}
G_2(z)=\sum_{n=1}^{k+1}c_{2\,n}\,z^{2\,n},
\end{equation}
where $c_{2\,k+2}$ is positive. Defining
\begin{align}\label{35}
\mu:=&(k+1)\,A\,c_{2\,k+2},\nonumber\\
\beta_n:=&\frac{k+1-n}{k+1}\,\frac{c_{2\,k+2-2\,n}}{c_{2\,k+2}},
\end{align}
and using (\ref{14}) and (\ref{15}), one finds
\begin{align}\label{36}
\rho_\mathrm{w}(z)=&\frac{\mu}{\pi}\,\sqrt{a^2-z^2}\,
\sum_{q=0}^k\sum_{n=0}^{k-q}\gamma_n\,\beta_{k-n-q}\,z^{2\,q},\nonumber\\
=&:\frac{\mu}{\pi}\,\sqrt{a^2-z^2}\,
\sum_{q=0}^k\delta_q\,z^{2\,q},
\end{align}
and
\begin{equation}\label{37}
\mu\,\sum_{n=1}^{k+1}\gamma_n\,\beta_{k+1-n}=1,
\end{equation}
respectively, where
\begin{equation}\label{38}
\gamma_n:=\frac{(2\,n-1)!!}{2^n\,n!}\,a^{2\,n}.
\end{equation}

The aim is to tune the coupling constants so that
\begin{equation}\label{39}
\left.\sum_{q=0}^k\sum_{n=0}^{k-q}\gamma_n\,\beta_{k-n-q}\,
z^{2\,q}\,\right|_{a=a_0}=z^{2\,k}.
\end{equation}
This leads to
\begin{equation}\label{40}
\left.\sum_{n=0}^p\gamma_n\,\beta_{p-n}\,\right|_{a=a_0}=\delta_{p,0},
\qquad p\leq k.
\end{equation}
To solve this, one notices that
\begin{equation}\label{41}
\sum_{n=0}^\infty\gamma_n s^n=(1-a^2\,s)^{-1/2}.
\end{equation}
It is seen that if one defines $\tilde\beta_n$ such that
\begin{equation}\label{42}
\sum_{n=0}^\infty\tilde\beta_n s^n=(1-a_0^2\,s)^{1/2},
\end{equation}
then
\begin{equation}\label{43}
\left.\sum_{n=0}^p\gamma_n\,\tilde\beta_{p-n}\,\right|_{a=a_0}=\delta_{p\,0}.
\end{equation}
So $\beta_n$ is the same as $\tilde\beta_n$, for $n\leq k$:
\begin{equation}\label{44}
\beta_n=-\frac{(2\,n-3)!!}{2^n\,n!}\,a_0^{2\,n},\qquad n\leq k.
\end{equation}
It is seen that $\beta_0$ is positive (in fact one), while other
$\beta_n$'s are negative. Now consider the coefficient of
$z^{2\,q}$ in the summation (\ref{36}). One has
\begin{equation}\label{45}
\delta_q=\gamma_{k-q}\,\left(1+\sum_{n=0}^{k-q-1}
\frac{\gamma_n}{\gamma_{k-q}}\,\beta_{k-n-q}\right).
\end{equation}
It is seen that for $q<k$, the derivative of the parenthesis with
respect to $a$ is positive. (Each term in the summation is a
negative constant times a negative power of $a$). Knowing that it
vanishes at $a=a_0$, one deduces that $\delta_q$ is positive for
$a>a_0$, negative for $a<a_0$, and there is a value $a_q$ less
than $a_0$, so that the derivative of $\delta_q$ is positive for
$a>a_q$. One then concludes that $\rho_\mathrm{w}$ is nonegative
for $a>a_0$, and $\rho_\mathrm{w}(0)$ is negative for $a<a_0$.
Also, the derivative of $\rho_\mathrm{w}(0)$ with respect to $a$
is positive for $a\geq a_0$. One can summarize these like
\begin{align}\label{46}
\rho_\mathrm{w}(z)>&0,\qquad a>a_0,\\ \label{47}
\rho_\mathrm{w}(0)<&0,\qquad a<a_0,\\ \label{48}
\left.\frac{\partial\rho_\mathrm{w}(0)}{\partial a}\right|_{a\geq
a_0}>&0.
\end{align}
Next consider (\ref{37}). The aim is to prove that with the choice
(\ref{44}), to every $a$ larger than $a_0$ there corresponds a
positive $\mu$, and for $a\geq a_0$ the derivative of $a$ with
respect to $\mu$ is negative. One can rewrite (\ref{37}) as
\begin{align}\label{49}
\frac{1}{\mu}=&-\tilde\beta_{k+1}+
\sum_{n=0}^{k+1}\gamma_n\,\tilde\beta_{k+1-n},\nonumber\\
=&-\tilde\beta_{k+1}+\delta_{-1}.
\end{align}
So, repeating exactly the same arguments used to arrive at
(\ref{46}) to (\ref{48}), one concludes for that $a\geq a_0$ the
right-hand side of (\ref{49}) is positive and the derivative of
the right-hand side of (\ref{49}) with respect to $a$ is positive:
\begin{align}\label{50}
\mu(a)>&0,\qquad a\geq a_0,\\ \label{51}
\left.\frac{\partial\mu}{\partial a}\right|_{a\geq a_0}<&0,\qquad
a\geq a_0.
\end{align}
The final picture is the following. Increasing $\mu$ from zero,
$a$ decreases from infinity, so that at $\mu=\mu_0$ one arrives at
$a=a_0$:
\begin{equation}\label{52}
\mu_0:=\frac{2^{k+1}\,(k+1)!}{(2\,k-1)!!\,a_0^{2\,k+2}}.
\end{equation}
For $\mu<\mu_0$, the density is nonnegative, while as $\mu$
exceeds $\mu_0$ the density becomes negative at $z=0$. (\ref{48})
and (\ref{51}) show that the derivative of $\alpha$ with respect
to $\mu$ is nonvanishing at $\mu=\mu_0$. Hence,
\begin{equation}\label{53}
l=1.
\end{equation}
To find the value of $m$, one notes that at the transition point
the first nonvanishing derivative of $\rho_\mathrm{w}$ at the
origin is the $(2\,k)$'th derivative. So,
\begin{equation}\label{54}
m=k.
\end{equation}
From these, one finds that the order of the transition is
$[2+(1/k)]$.

There remains one point to take into account, which is to make
sure that for $\mu\leq\mu_0$ the density does not exceed one. To
address this, one notices that the transformation
\begin{align}\label{55}
a\to &\, \sigma\,a,\nonumber\\
a_0\to &\, \sigma\,a_0,\nonumber\\
\mu\to &\, \sigma^{-2\,k-2}\,\mu,\nonumber\\
z\to &\, \sigma\,z,\nonumber\\
\rho\to &\, \sigma^{-1}\,\rho,
\end{align}
where $\sigma$ is an arbitrary positive constant, leaves all of
the relations intact. So, using a sufficiently large $\sigma$
ensures that the DK transition does not occur before the
transition from the weak regime to the gaped regime.\\
\\
\textbf{Acknowledgement}:  M.A. would like to thank the research
council of the University of Tehran for partial financial support.

\end{document}